\documentclass[preprint]{revtex4}
\usepackage{graphicx}
\usepackage{amssymb}
\usepackage{amsmath}
\usepackage{epsfig}
\usepackage{stmaryrd}

\begin{document}

\title{Nanomechanical properties of few-layer graphene membranes}
\author {M. Poot and H. S. J. van der Zant\footnote{h.s.j.vanderzant@tudelft.nl}}
\affiliation{Kavli Institute of Nanoscience, Delft University of Technology, Lorentzweg 1, 2628 CJ Delft, The Netherlands} \date{\today}
\vspace{1cm}
\begin{abstract}
We have measured the mechanical properties of few-layer graphene and graphite flakes that are suspended over circular holes. The spatial profile of the flake's spring constant is measured with an atomic force microscope. The bending rigidity of and the tension in the membranes are extracted by fitting a continuum model to the data. For flakes down to eight graphene layers, both parameters show a strong thickness-dependence. We predict fundamental resonance frequencies of these nanodrums in the GHz range based on the measured bending rigidity and tension.
\end{abstract}

\maketitle
\newpage

Graphene, a single layer of graphite, has recently been contacted with electrodes \cite{Novoselov_first} and its unique electronic properties are being measured \cite{Novoselov_AQHE, Zhang, Novoselov_RT, Heersche}. By suspending graphene, membranes of only one atom thick are obtained \cite{Meyer}, which may have interesting applications, such as pressure sensors or gas detectors \cite{Schedin} or they can be used to build mechanical resonators \cite{Bunch}. In this Letter, we present a method to obtain the bending rigidity of and the tension in ultra-thin membranes by fitting the spatial profile of the compliance. We applied this method to suspended multi-layer graphene. Over almost four decades, the bending rigidity closely follows the thickness-dependence for graphite, calculated using continuum mechanics.

Samples are made from doped silicon wafers with 285 nm silicon oxide on top, in which circular holes are etched with buffered hydrofluoric acid using resist masks\cite{Witkamp}. Graphite grains are put on adhesive tape, cleaved and the tape is pressed against the substrate \cite{Novoselov_first}. This way, graphitic flakes with varying dimensions are left on the surface, covering some of the holes completely as Fig. 1a shows.

The elastic properties of more than 50 flakes with thicknesses varying from 2.4 nm to 33 nm (8 to 100 layers) are extracted from ensembles of force-distance curves, measured with an atomic force microscope (AFM) under ambient conditions: The deflection of the AFM tip, $z_\mathrm{tip}$, is measured while lowering the tip onto the sample over a distance $z_\mathrm{piezo}$, as illustrated in Fig. 1b and c. The deflection of the flake $u$ is due to the applied force: $F =k_{\mathrm{tip}}z_\mathrm{tip}$, where $k_{\mathrm{tip}}$ is the spring constant of the AFM tip \cite{ktip}. The (negative) slope of the force-distance curve $s = -\mathrm{d}z_\mathrm{tip}/\mathrm{d}z_\mathrm{piezo}$, is used to extract the local compliance of the flake $k_\mathrm{f}^{-1} = \mathrm{d}u/\mathrm{d}F_\mathrm{tip} = k_\mathrm{tip}^{-1}(s^{-1}-1)$. However, knowing the complicance at a single point is not enough to extract all mechanical properties of a membrane \cite{Norouzi}. Therefore, multiple force-distance curves are recorded while scanning in a rectangular grid over the sample to construct a map of the local compliance. This is the so-called force-volume method \cite{Radmacher}. 

Fig. 1c shows two individual force-distance curves out of a set of $64\times64$ curves. The lower curve was taken on an unsuspended part of the flake, while the other was taken on a suspended part. The deflection of the flake results a lower slope in the latter case. The curves are linear (apart from the small region where the tip is almost touching the flake) for deflections up to a quarter of the thickness, i.e. the deflection of the flake is proportional to the applied force. Whenever a non-linear force-distance curve was observed, the applied force was reduced significantly to ensure that the measurements were done in the linear regime. Note that with the force modulation technique \cite{Maivald}, this check is not possible, as only the slope $s$ is measured. Another advantage of the force-volume method is the absence of lateral forces on the flake while scanning, which might strain or even damage the flakes. 

Fig. 2a shows a map of the local compliances extracted from a force-volume measurement. In this plot, different regions can be distinguished: In the upper left corner, the tip presses against the hard silicon oxide and the compliance vanishes. The edge of the flake appears as a line of high compliance, because the tip slides along the edge when pressing. The light blue color indicates that even a supported part of the flake has a non-zero compliance, i.e., it is indentable. This is not surprising when the low Young's modulus $E_\perp$ = 37 GPa of graphite \cite{Blakslee} for stress perpendicular to the graphene planes is considered. We found no clear correlation between the indentability and the thickness of the flake, probably due to the different tip geometries in the measurements. Although not visible in the height image, the hole appears as a circular region with high compliance. At the center of the hole, the flake is more easily deflected than at the edge, as expected. 

To find the bending rigidity of and tension in the membranes, a continuum model \cite{continuum} for the induced deflection is developed, which is fitted to the experimental data. The AFM tip is modeled as a point force, as it's radius of curvature (of the order of 10 nm) is much smaller than the radius of the hole $R$. This differs from studies on lipid bilayer membranes, where the hole diameter is of the same order as the radius of the tip\cite{Steltenkamp}. The force applied at $(r_\mathrm{0},\theta_\mathrm{0})$ is opposed by the bending rigidity $D$ and the tension $T$, which we assume to be isotropic \cite{tension}, i.e. the flake is equally stretched in both horizontal directions. The equation for deflections that are small compared to the thickness $h$ is \cite{Norouzi, LL}:
\begin{equation}
\big(D\nabla^4 - T \nabla^2\big) u(r, \theta; r_\mathrm{0},
\theta_\mathrm{0}) = \frac{F_\mathrm{tip}}{r}\delta(r-r_\mathrm{0},\theta-\theta_\mathrm{0}),
\label{eq_deflection}
\end{equation}
which is solved analytically for a flake that is clamped at the edge of a circular hole (i.e., $u(R) = 0$ and $\mathrm{d}u/\mathrm{d}r(R) = 0$). A calculated deflection profile is shown in Fig. 2b. The AFM measures the deflection at the point where the force is applied: $u(r_\mathrm{0},\theta_\mathrm{0}; r_\mathrm{0},\theta_\mathrm{0})$, which is proportional to the applied force, with the local compliance as the proportionality factor, and is independent of $\theta_0$. Large deflections introduce terms proportional to $u^3$ in Eq. 1\cite{LL}, which would result in non-linear force-distance curves. As the measured force-distance curves are linear, no higher order terms have to be included in the model. By varying the location of the applied force $r_0$, the compliance profile can be calculated. It depends on three fitting parameters: the bending rigidity $D$, the tension $T$ and the radius of the hole $R$. 

As shown in Fig. 2, good fits (solid lines) are obtained with this model. When the measurement is repeated on the same hole, the fit parameters differ less than a few percent. The hole radius from the fit is in agreement with height profiles and scanning electron microscopy. Fig. 2c shows a profile that is rounded at the edge of the hole. This is reproduced by a fit, where the compliance is primarily due to the bending rigidity. The profile in Fig. 2d is sharper at the edge, which can be fitted well with a large tension. The question whether a flake is tension or rigidity dominated can only be answered with mechanical measurements, as the height maps do not show any difference. The extracted bending rigidity of every flake is plotted in Fig. 3a against its thickness. It increases strongly with the thickness, while at the same time the spread increases. Measurements on a flake suspended over holes with different diameters confirm that the bending rigidity does not depend on the hole size, but that it is an intrinsic property of the flake. 

The bending rigidity of bulk graphite can be calculated using continuum mechanics. Graphite is highly anisotropic, but the in-plane mechanical properties are isotropic and can be described by the in-plane Young's modulus $E_\boxempty =$ 0.92 TPa and the in-plane Poisson's ratio $\nu_\boxempty$ = 0.16 \cite{Blakslee}. The bending rigidity for deflections perpendicular to the graphene planes is obtained by generalizing the analysis in \cite{LL} to the anisotropic case, giving: $D = E_\boxempty h^3/12(1-v^2_\boxempty)$. The gray line in Fig. 3a shows this relation; Over the entire range, most values for the bending rigidity are close to this curve. Only flakes thicker than about 10 nm may have a smaller bending rigidity. A possible explanation for this deviation is the presence of stacking defects in the flakes: the bending rigidity is no longer proportional to $h^3$, but in a first approximation to the sum of the cubes of the thickness of each part separated by the defects, resulting in a smaller bending rigidity. This is also consistent with the fact that the spread in the obtained values grows with increasing thickness. For flakes with $h < 10$ nm, the data points are close to the drawn line in Fig. 3a, which would imply the absence of stacking faults in thin flakes. 

The tension varies from flake to flake and its thickness-dependence is shown in Fig. 3b. The tension is larger for thicker flakes, possibly saturating at 20 N/m, but more measurements are needed to confirm this. Measurements on different holes underneath the same flake give similar values for the tension, so the tension is uniform throughout the flake. Most likely, the tension is induced during the deposition process \cite{Bunch}.

Knowing the experimental values for the bending rigidity and tension, other mechanical properties can be calculated. As an example Fig. 3c shows the expected eigenfrequencies\cite{Wah} of the flakes, calculated with the measured values of the bending rigidity and tension. The frequency increases with increasing thickness. For holes with $R$ = 0.54 $\mu$m, the frequencies are slightly below 1 GHz, while for smaller holes ($R$ = 84 nm), the frequency can be over 10 GHz. These high resonance frequencies make our nanodrums ideal components for nanomechanical devices. 

In conclusion, we have shown that an AFM measurement of the compliance profile of a suspended membrane yields important information on it's mechanical properties. This technique is not limited to multi-layer graphene flakes, but can be applied to membranes of any kind.

We thank J\"orgen Konings and Abdulaziz Karimkhodjaev for their help with the measurements and Alberto Morpurgo, Samir Etaki, Jari Kinaret and Andreas Isacsson for discussions. Financial support is obtained from the Dutch organizations FOM, NWO (VICI-grant) and NanoNed.

\newpage

\begin{figure}[tbp]
	\centering		
	\includegraphics[width=8.5cm]{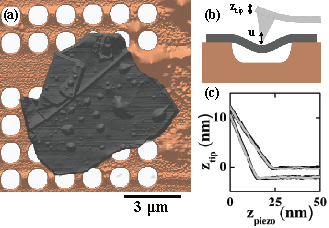}
	\caption{(a) An AFM height image of a suspended flake ($\sim$69 layers). (b) Schematic overview of the method used to determine the local compliance of the flake. (c) Two linear force-distance curves (offset for clarity) taken on the flake shown in (a). The approaching (gray) and retracting (black) parts of the curves lie on top of each other. The bottom curve is taken on an unsuspended part of the flake, while the top curve is taken on a suspended part.}
	\label{fig1}
\end{figure}

\begin{figure}[tbp]
	\centering		
	\includegraphics[width=8.5cm]{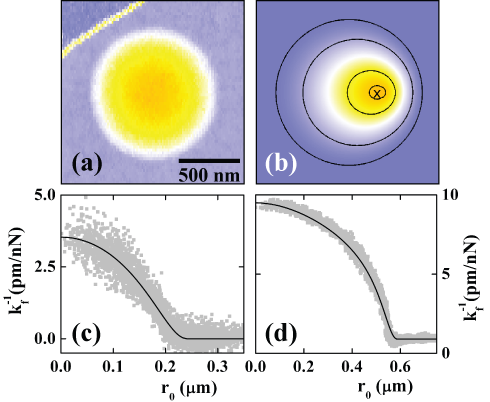}
	
	\caption{(a) Colormap of the compliance of a flake with $h$ = 23 nm, extracted from a force-volume measurement ($64\times64$ force-distance curves) . The compliance ranges from 0 (blue) to $9.7\cdot10^{-3}$ m/N (orange). (b) The calculated deflection in the absence of tension for a force applied at the position of the cross. (c) The measured radial profile of the compliance (symbols) of a 15 nm thick flake and the fit by the model (solid line). (d) The radial profile of the data shown in (a).}
	\label{fig2}
\end{figure}

\begin{figure}[tbp]
	\centering		
	\includegraphics[width=8.5cm]{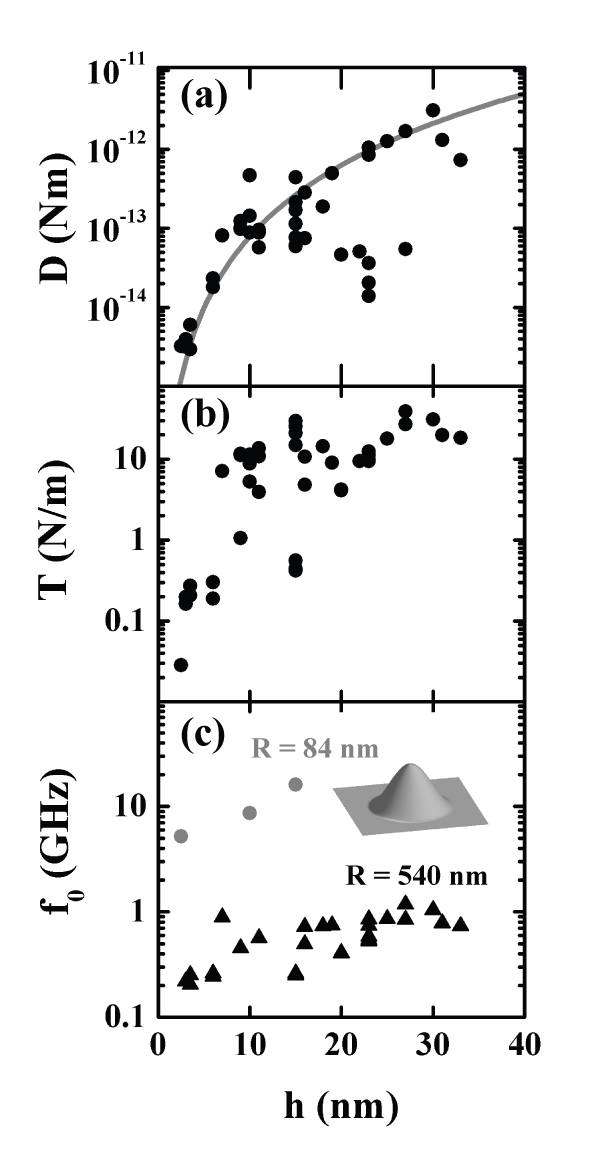}
	\caption{Thickness-dependence of the mechanical properties extracted from the fits. (a) The bending rigidity $D$ (symbols) and the continuum relation (gray line). (b) The tension in the flake $T$. (c) The frequency of the fundamental mode $f_0$ calculated with the measured values for $D$ and $T$ for two different hole sizes. The inset shows the displacement profile of this mode.}
  \label{fig3}
\end{figure}


\begin{thebibliography}{99}
\bibitem{ktip} Commercially available AFM tips with nominal spring constants $k_\mathrm{tip}$ = 2 or 42 N/m are used. The spring constant is calibrated using the thermal noise method and the deflection sensitivity is obtained by taking an ensemble of force-distance curves on the silicon oxide substrate.
\bibitem{continuum} Even when the continuum approximation fails along the z-direction, it can still be valid in the horizontal direction, as long as the z-dependence of the induced deformations is not considered. Therefore, the bending rigidity is not expressed in the elastic constants, but it is treated as a fit parameter.
\bibitem{tension} We have calculated the first order correction to $u$ for an anisotropic tension: The flake is stiffer than average along the principal direction with the largest tension and weaker along the other. The lines of constant compliance are ellipsoidal instead of circular. As this has not been observed in the measurements, the assumption of isotropic tension is allowed.
\end{thebibliography}
\end{document}